\documentclass[a4paper,onecolumn,12pt,assc]{IEEEtranASSC}
\usepackage{cite}
\usepackage[cmex10]{amsmath}
\usepackage{color}
\usepackage[dvips]{graphicx}
\usepackage[tight,footnotesize]{subfigure}

\begin{document}
\title{A dynamical test for terrestrial planets in the habitable zone of HD~204313}

\author{\IEEEauthorblockN{Elodie Thilliez\IEEEauthorrefmark{1}, Lea Jouvin\IEEEauthorrefmark{2}, Sarah T. Maddison\IEEEauthorrefmark{1} and Jonathan Horner\IEEEauthorrefmark{3}\IEEEauthorrefmark{4}}

\IEEEauthorblockA{\IEEEauthorrefmark{1}
Centre for Astrophysics \& Supercomputing, Swinburne University of Technology, Hawthorn, VIC, 3122, Australia}

\IEEEauthorblockA{\IEEEauthorrefmark{2}
Universite Paris-Sud 11, 91400 Orsay, France}

\IEEEauthorblockA{\IEEEauthorrefmark{3}
School of Physics, University of New South Wales, Sydney, NSW, 2052, Australia}

\IEEEauthorblockA{\IEEEauthorrefmark{4}
Australian Centre for Astrobiology, University of New South Wales, Sydney, NSW 2052, Australia}}

\maketitle

{\color{red} Accepted for publication in the peer-reviewed proceedings of the 13th Australian Space Science Conference, held at the University of New South Wales, Sept. 30th $-$ Oct. 2nd 2013.}

\begin{abstract}
With improvements in exoplanet detection techniques, the number of multiple planet systems discovered is increasing, while the detection of potentially habitable Earth-mass planets remains complicated and thus requires new search strategies. Dynamical studies of known multiple planet systems are therefore a vital tool in the search for stable and habitable planet candidates.

Here, we present a dynamical study of the three-planet system HD~204313 to determine whether it could harbour an Earth-like planet within its habitable zone for a sufficient time to develop life. We found two semi-stable regions in the system, but neither prove stable for long enough for a terrestrial planet to develop life.  Our investigations suggest that overlapping weak and high order resonances may be responsible for these semi-stable regions. This study established a framework for a larger project that will study the dynamical stability of the habitable zone of multiple planet systems, providing a list of interesting targets for future habitable low-mass planet searches.
\end{abstract}


\section*{Introduction}
Since the discovery of the first exoplanet, more than a thousand exoplanets have been confirmed, including 170 multiple planet systems\footnote{Current exoplanet tally taken from the Extrasolar Planets Encyclopaedia, httest particle://www.exoplanet.eu, on the 11th November 2013}. With improvements in exoplanet detection techniques, new discoveries are expected and will continue to enhance our understanding of planetary systems and the planet formation process. However the detection of habitable Earth-mass planets remains complicated, and only 12 super-Earth planets are currently known to orbit within the habitable zone of their host star~\cite{Bor13}, while 35 additional Kepler candidates\footnote{Current estimation from the Habitable Exoplanets Catalog, http://phl.upr.edu, on the 11th November 2013} are still waiting to be confirmed.

The habitable zone (HZ) is defined as the region around a star where liquid water could be present on the surface of rocky planets~\cite{Kas93}. The presence of liquid water on the surface depends on a broad range of parameters, such as the stellar incident energy, spectral distribution, radiative properties of the planetary atmosphere (absorption, diffusion, emission, clouds) and the reflectivity of its surface (albedo)~\cite{Kas93}. Earth is the only habitable planet of which we currently know, and it appears that life took $\sim$ 1 Gyr to develop a significant atmospheric signature that would be detectable~\cite{Chy93}. Using the Earth as a baseline, we will therefore work on the assumption that a planet would need to survive for at least 1 Gyr within a star's habitable zone to detect any life there. This therefore requires additional constraints to be placed on the orbit of the planet for it to be deemed detectably habitable - not only must it currently orbit within the habitable zone, but it must also maintain an orbit which is simultaneously dynamically stable and confined within the HZ (i.e. a small eccentricity) for this entire duration.

Dynamical studies thus play a crucial role in the hunt for habitable exoplanets, and the evaluation of the long-term stability of the HZ of known multiple exoplanet systems can help focus future observations. There are a number of examples in the literature of multiple planet systems which turn out to be dynamically unstable on short timescales. For example, the two planet system orbiting the catacylsmic variable HU Aquarii announced by \cite{Qia11} was shown by \cite{Hor11} and \cite{Wit12} to be unstable on timescales of order $10^{4}$ years, strongly suggesting that the multiplanet system is not real or that the orbital parameters must be different from those reported in the literature. While numerical simulations have been commonly used and are necessary to understand the dynamics of multiple systems (e.g.~\cite{San07,Hin08,Hor11}), analytical stability criteria tools can be used to provide a quick assessment of the stable regions of a multiplanet system (e.g~\cite{Dec13,Giu13}).

We aim to establish a new framework to systematically investigate the stability of HZs of multiple planetary systems. Firstly, the HZ of such systems must be located using atmospheric model criteria, and then a numerical search for dynamically stable regions within the HZ can be conducted. This requires the integration of the orbits of massless test particles randomly distributed inside the HZ. Once stable regions (where test particles survive at least 1 Gyr) are identified, we can then explore their long-term stability by (i) investigating their resonance state, and (ii) producing maps of their stability as a function of orbital elements by varying the stable test particles initial conditions around the stable known initial parameters. Finally, to ensure stability of a potential Earth-mass planets in these stable and habitable zones, numerical simulations must be run by substituting the stable test particles with Earth-mass planets. 

In this paper we present a study of the dynamical stability of test particles in the habitable zone of the star HD~204313. Using the N-body integrator \textsc{SWIFT}~\cite{Lev94}, we first isolate regions of stability within the habitable zone of HD~204313 using massless test particles. We then numerically investigate the long-term stability of test particles in the identified dynamically stable regions of the HZ and compare these locations with the stable regions derived using the analytical stability criteria of~\cite{Giu13}. The paper is organized  as follows: first, we describe the HD~204313 system, and the location of its habitable zone, in the following section. We then present the analytical and numerical methods used to investigate the stability of the system's habitable zone. We then report our findings in section 3 before discussing the potential origin of two identified semi-stable zones in section 4.

\section*{The HD~204313 System and its Habitable Zone}
HD~204313 is a metal-rich Sun-like star of spectral type G5~\cite{Kha09}, with a mass of 1.02 $M_{\odot}$ and age of 7.2 Gyrs~\cite{Cas11}. It has been the target of multiple radial velocity surveys and was observed by~\cite{Rob12} with  the 2.7 m Smith telescope over 8 years to obtain Radial Velocity measurements and a stellar spectrum. Using the MOOG local thermodynamic equilibrium line analysis and spectral analysis program~\cite{Sne73}, they determined the effective temperature, $T_{eff}$ and by matching the abundances of the Fe I and Fe II lines, they determined the metallicity of HD~204313, [Fe/H]. The star's distance and absolute magnitude, $m_{V}$, were also derived by~\cite{Rob12} and all their stellar properties are summarized in Table \ref{Table1}. \\
\begin{table}[!hb]
\renewcommand{\arraystretch}{1.0}
\caption{Stellar properties for HD~204313 from~\cite{Rob12}.}
\label{Table1}
\centering
\begin{tabular}{|c||c|}
\hline
\hline
Property & Value\\
\hline
Spectral type$^{a}$ & G5 V \\
Age$^{b}$ & 7.2 Gyrs \\
Mass$^{b}$ & 1.02~$M_{\odot}$ \\
Distance & 47.0 $\pm$ 0.3~pc\\
$T_{eff}$ & 5760 $\pm$ 100~K \\
$m_{V}$ & 4.63 $\pm$ 0.03 \\
$[$Fe/H$]$ & 0.24 $\pm$ 0.06\\
\hline
\end{tabular}
\\
\small{$^{a}$ From \cite{Kha09}\\
$^{b}$ From \cite{Cas11}, maximum likelihood estimate.}
\end{table}

This system is known to harbour at least three exoplanets: \cite{Seg10} presented the discovery of HD~204313 b, a super-Jupiter planet ($M sin(i)\sim~$3.5 $M_{J}$) with a period $P \sim$ 5 years with the CORALIE Echelle spectrograph, while \cite{Mayo11} revealed an interior ($P \sim$ 35 days) Neptune-mass planet HD~204313 c with the HARPS spectrograph. \cite{Rob12} simultaneously fitted the data from the CORALIE Echelle spectrograph in conjunction with their observations from the Harlan J. Smith Telescope at McDonald Observatory. The residuals from their fully generalised Lomb-Scargle periodogram for a one-planet model (HD~204313 b) showed significant peaks, notably around $P$ $\sim$~ 2700 days. By fitting an additional planet at this period, their fitting routine converged for an additional Jupiter-mass planet: HD~204313 d with mass $Msin(i)$ $\sim~$1.68 $M_{J}$, and a period of 2831 days. They did not include the planet HD~204313 c in their study since its radial velocity amplitude as detected by \cite{Mayo11} was below the sensitivity limit of their data, but that planet could be responsible for the remaining scatter in their two planet fit. We will be using the orbital parameters derived by \cite{Rob12} for the two planets, HD~204313 b and d, in our study -- see Table \ref{Table2}. Their best-fit orbits reveal that HD~204313 b and d are trapped in a mutual 3:2 mean-motion resonance (MMR). By performing further dynamical analysis, the authors found that outside this MMR, there is no stable configuration for these two planets, and that fine tuning of orbital elements of the two planets was required in order for this system to survive. Moreover they confirmed that adding the inner planet, HD~204313 c, to their dynamical analysis does not affect the behaviour of HD~204313 b and d.\\
\begin{table}[!hb]
\renewcommand{\arraystretch}{1.0}
\caption{Orbital parameters of HD~204313 planets from \cite{Rob12}}
\label{Table2}
\centering
\begin{tabular}{|c||c|c|c|}
\hline
\hline
Elements & HD~204313$^{a}$ c & HD~204313 b & HD~204313 d\\
\hline
$M\sin(i)$ ($M_{J}$) & 0.054 $\pm$ 0.005 & 3.55 $\pm$ 0.2 & 1.68 $\pm$ 0.3\\
$a$ (AU) & 0.2103 $\pm$ 0.0035 & 3.04 $\pm$ 0.06 & 3.93 $\pm$ 0.14\\
$e$ & 0.17 $\pm$ 0.09 & 0.23 $\pm$ 0.04 & 0.28 $\pm$ 0.09 \\
$P$ (days) & 34.88 $\pm$ 0.03 & 1920 $\pm$ 25 & 2830 $\pm$ 150 \\
$M$ $(^{o}$) & & 300 $\pm$ 0.4 & 137 $\pm$ 2 \\
$\omega$ $(^{o}$) & & 298 $\pm$ 6 & 247 $\pm$ 16 \\
\hline
\end{tabular}
\\
\small{$^{a}$ From \cite{Mayo11}.}
\end{table}

As discussed earlier, the location of the HZ around a given star depends both on the nature of the star itself (parameters such as its luminosity and temperature) and the atmosphere of the planet in question (absorption, diffusion, albedo, emission efficiency and cloud coverage)~\cite{Kas93}. Thus the location of the HZ is expected to migrate outward as the star evolves on the main sequence and its effective temperature and luminosity increase. Therefore the time available for a planet to develop life in its HZ depends on (i) the width of this zone and (ii) the lifetime of a star on its main sequence (which depends of the stellar mass and metallicity)~\cite{Und03}. There are several ways to define the inner and outer boundary of the HZ (as discussed in~\cite{Jon05}, \cite{Jon06},\cite{Und03} and \cite{Men03}). We select the inner boundary as the maximum distance from the star at which a `runaway greenhouse effect' would lead to the evaporation of all surface water, while for the outer boundary we chose the `maximum greenhouse effect', which defines the distance at which a cloud-free CO$_{2}$ atmosphere could maintain a surface temperature of 273~K.
  
To calculate the distance of these boundaries from the star, we followed the work of~\cite{Kop13}, who express the effective flux received at the HZ boundaries, $S_{eff}$, as a quadratic function of the difference between the effective temperature of the star and that of the Sun, $T_{eff}-T_{\odot}$. The slope of this function is given by coefficients that depend on the boundary criteria of the HZ, with all coefficients for each boundary given in \cite{Kop13}. We selected the coefficients for a `runaway greenhouse effect' as inner boundary and the coefficients for a `maximum greenhouse effect' as an outer boundary. The ratio between the luminosity of the star and our Sun , $L/L_{\odot}$, is then deduced from its apparent visual magnitude, $m_{V}$, and from its distance to our Sun, $D$, in pc. Finally, the corresponding distance, $d_{\rm HZ}$, of the limits of the HZ is given by:
\begin{equation}
\centering
d_{\rm HZ}=\left( \frac{1}{S_{eff}}\frac{L}{L_{\odot}} \right)^{1/2}
\end{equation}
Using the stellar parameters from Table~\ref{Table1}, we find the HZ boundaries for HD~204313 at:
\begin{center}
 $HZ_{inner}$ (runaway greenhouse) = 1.1 AU \\
 $HZ_{outer}$ (maximum greenhouse) = 1.9 AU
\end{center}

\section*{Numerical and Analytic methods}
We expand on the work of~\cite{Rob12}, who conducted a dynamical stability study of the HD~204313 system to constrain the orbital parameters of the two outer planets using the N-body dynamics package \textsc{Mercury}~\cite{Cham99}. Because of its negligible mass compared to the other two planets and its short period, HD~204313~c was found to have little impact on the system's dynamics: its Hill radius, which defines its zone of gravitational influence, is $\sim$ 0.005~AU which means that it is dynamically well separated from the two outer planets and the HZ, and therefore does not significantly perturb them. As a result, those authors did not include that planet in their long-term simluations. In their study, the orbital parameters of HD~204313~b were kept fixed, while those of HD~204313~d were varied: the eccentricity, $e$, semi-major axis, $a$, longitude of periastron, $\omega$, and mean anomaly, $M$, of HD~204313 d were systematically varied by $\pm$ 3$\sigma$ around their best fit values, with simulations run for a maximum $10^{8}$ years. The orbits of the two giant planets are assumed to be coplanar. They derived two stability maps displaying the mean dynamical lifetime of the HD~204313~b and d planetary system as (i) a function of the initial semi-major axis and eccentricity of planet d, and (ii) as a function of the initial semimajor axis and longitude of periastron of planet d -- see their Figure 5~\cite{Rob12}.

We used the \textsc{Swift} integration software package to numerically integrate the orbits of HD~204313~b and HD~204313~d over a similar duration ($10^{8}$ years). We used the N-body symplectic RMVS integrator of \textsc{Swift}, a well-tested integrator first computed by~\cite{Lev94} which conserves the Hamiltonian during the integration and integrates close encounters between planets and test particles. Their goal was to model the orbit of short-period comets under the gravitational influence of all the Solar system's planets except Mercury. To ensure that \textsc{Swift} gives similar results to \textsc{Mercury}, we repeated a subsample of the simulations of~\cite{Rob12} with \textsc{Swift}, selecting 13 simulations along the edge of the stable zone in both $a,e$ and $a,\omega$ space -- see Table~\ref{Table3}. 
\begin{table}[!hb]
\renewcommand{\arraystretch}{1.0}
\caption{Initial orbital parameters of HD~204313~d for the 13 simulations taken from~\cite{Rob12}}
\label{Table3}
\centering
\begin{tabular}{|c||c|c|c|c|c|c|c|c|c|c|c|c|c|}
\hline
\hline
 & 1 & 2 & 3 & 4 & 5 & 6 & 7 & 8 & $9^{a}$ & 10 & 11 & 12 & $13^{a}$ \\
\hline
$a$ (AU) & 3.90 & 3.93 & 3.93 & 3.98 & 3.98 & 3.98 & 3.98 & 3.98 & 4.01 & 4.01 & 4.01 & 4.01 & 4.07 \\
$e$ & 0.41 & 0.12 & 0.3 & 0.05 & 0.05 & 0.12 & 0.12 & 0.30 & 0.05 & 0.05 & 0.24 & 0.24 & 0.21 \\
$\omega$ ($^{o}$) & 292 & 256 & 283 & 238 & 247 & 247 & 256 & 292 & 238 & 247 & 283 & 292 & 274 \\
$M$($^{o}$) & 137 & 137 & 137 & 137 & 137 & 137 & 137 & 137 & 137 & 137 & 137 & 137 & 131 \\
\hline
\end{tabular}
\small{$^{a}$ Models 9 and 13 were found to be unstable.}
\end{table}
We found that 11/13 of the simulations were stable for $10^{8}$ years, with the two unstable models originating at the extreme edge of the stability zone of~\cite{Rob12} (see their Figure 5). Given that \textsc{Mercury} and \textsc{Swift} do not handle close encounters in exactly the same manner, it is not surprising to see some differences for critical cases. Whilst \textsc{Mercury} treats a close encounter between a planet and  test particle by switching from a sympletic integrator to a classic Bulirsch$-$Stoer integrator to ensure the precision set by the user is reached, \textsc{Swift} treats the close encounter using the same symplectic RMVS integrator but switching to the planet as the barycenter of the system instead of the star during the encounter.

Using these 11 stable models, we investigated the system's suitability for harboring additional stable bodies by randomly distributing 2000 massless test particles in the HZ of HD~204313 and integrating for $10^{8}$ years. The test particles had initial eccentricities $e_{TP}$ ranging from 0.0 to 0.3, inclinations $i_{TP}$ of 0 degrees, and random values of longitude of periastron, $\omega_{TP}$, and mean anomaly, $M_{TP}$. The outcome of these simulations allowed us to identify stable regions within the HZ. We then inspected the long-term stability of these zones by determining the resonance state of the stable particles, and produced maps of the test particle lifetimes as a function of their initial orbital elements. To create these maps, we ran additional simulations with test particle orbital elements ($a_{TP}, e_{TP}, \omega_{TP}, M_{TP}$) distributed over a specific range of values: we examined 51 values of $a_{TP}$ in a radius of 0.6~AU around the stable zone; 51 values of $e_{TP}$ between 0.0--0.3 for each $a_{TP}$ ; 19 values of $\omega_{TP}$ between 0.0--360$^{o}$ for each $e_{TP}$ value and 19 values of $M_{TP}$ between 0.0--360$^{o}$ for each $\omega_{TP}$ value. Thus more than $10^{6}$ test particles with unique initial conditions were generated to derive a high resolution stability map as a function of orbital elements inside the HZ.

By following this method, the numerical simulations can provide a complete picture of the stability state of test particles in the HZ. However, to ensure no dependence on exact initial conditions, this method requires a large number of simulations ($10^{6}$ in our experiment). Analytical criteria are another way to quickly appraise the location of stable regions. By generalizing the resonance overlap criterion for the onset of stochastic behaviour in the planar circular restricted three-body problem from~\cite{Wis80} to multiplanetary systems, \cite{Giu13} provide analytical expressions that can put constraints on the location of stable zones. Their criterion is based on that the fact that if the orbits of planets in a multiplanetary system cross and are not protected by some resonant mechanism, this could lead to the planets in the system colliding with one another, or being ejected from the system as a result of their mutual interactions. More precisely, around each planet $j$ exists a region of instability, $\delta_{j}$, where a close encounter with another body will perturb the dynamics enough to lead to chaotic diffusion of the eccentricity and semi-major axis, resulting in either collision or escape. The width of this region depends on the mass, $m$, eccentricity, $e$, and semi-major axis, $a$, of the planet, and on the mass of the encountering body. In our case, we will apply this criterion of the two planets HD~204313~b and HD~204313~d encountering an Earth-mass planet. Thus $\delta_{j}$ is given by~\cite{Giu13}:
\begin{equation}
\delta_{j}=1.57\times a(j) \times \left( \left( \frac{m(j)}{m_{star}} \right)^{2/7} + \left( \frac{m_{earth}}{m_{star}} \right)^{2/7} \right).
\end{equation}
In the simple case of a planet $j$ on a circular orbit, as long as no other bodies enter the region $a_{j}\pm \delta_{j}$ around the planet, the bodies will be stable with respect to planet $j$. For eccentric planets such as HD~204313 b and d, this region is extended to $q_{j}-\delta_{j}$ and $Q_{j}+\delta_{j}$ , where $q_{j}$ is the periastron and $Q_{j}$ the apastron of planet. We use the expressions derived by~\cite{Giu13} to assess the position of the interior and exterior limit of the unstable zone around HD~204313 b and d, and compare it to the stability zone of our test particles in the HZ from our numerical simulations.

\begin{figure}[t]
\centering
\begin{subfigure}
{\includegraphics[width=3.5in]{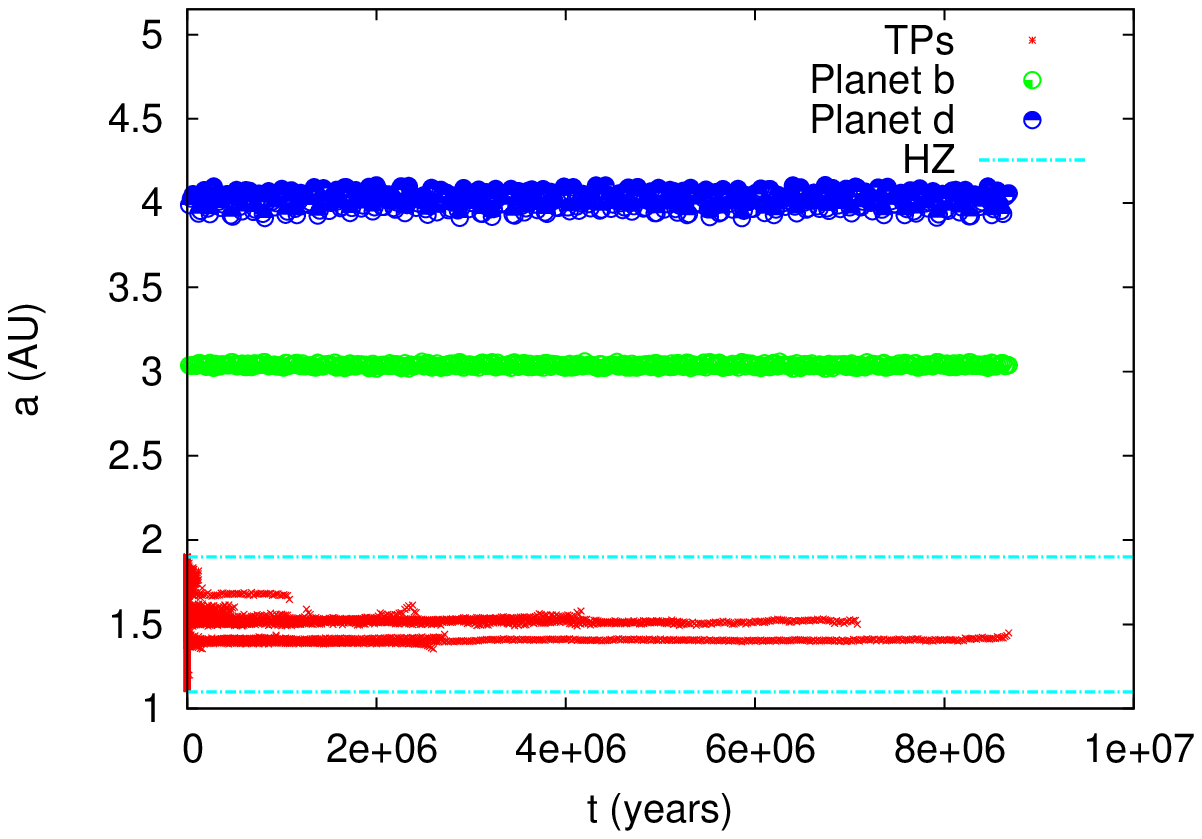}}
\caption{\small{Evolution of the semi-major axis, $a$, for the test particles and planets in model 4. The dotted lines represent the location of the habitable zone.}}
\label{fig:1} 
\end{subfigure}
\begin{subfigure}
{\includegraphics[width=3.5in]{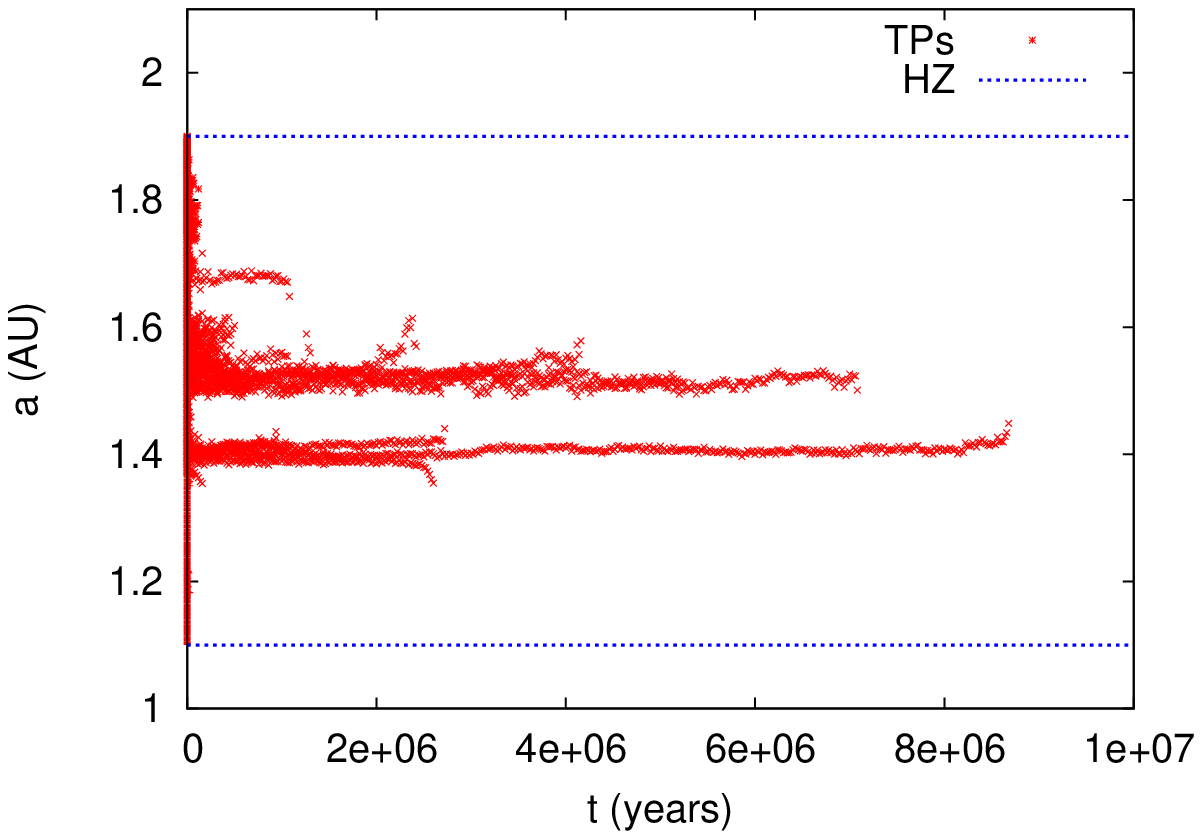}}
\caption{\small{Zoom of the HZ region of model 4, following the evolution of the semi-major axis, $a$, for the test particles. The dashed lines represent the location of the habitable zone.}} 
\label{fig:2}
\end{subfigure}                
\end{figure}
\section*{Results \& Discussion}
In each of our simulations, all of the test particles were removed from the HZ on timescales far shorter than the maximum possible run time ($10^{8}$ years). As a result, the integrations were terminated when the final test particle was removed - and so no simulation actually completed its full 100 million year run time. Figure~\ref{fig:1} shows the time evolution of the semi-major axis, $a$, for the test particles and the two planets for the longest lasting model in which the final test particle was ejected after a period of $\sim~9\times10^{6}$ years (model 4 in Table~\ref{Table3}). In all other models, the time at which the final test particle was removed was always significantly shorter. Figure~\ref{fig:2} zooms into the HZ region of model 4, in which it is apparent that all test particles in the HZ are rapidly removed by the gravitational influence of the two outer giant planets, aside from two areas located around $a\sim$~1.5 AU and $a\sim$~1.4 AU. We observed the same stable regions in each of our 11 simulations, and those regions were always located at the same distance from the central star.
\begin{figure}[t]
\centering
\begin{subfigure}
{\includegraphics[width=3.5in]{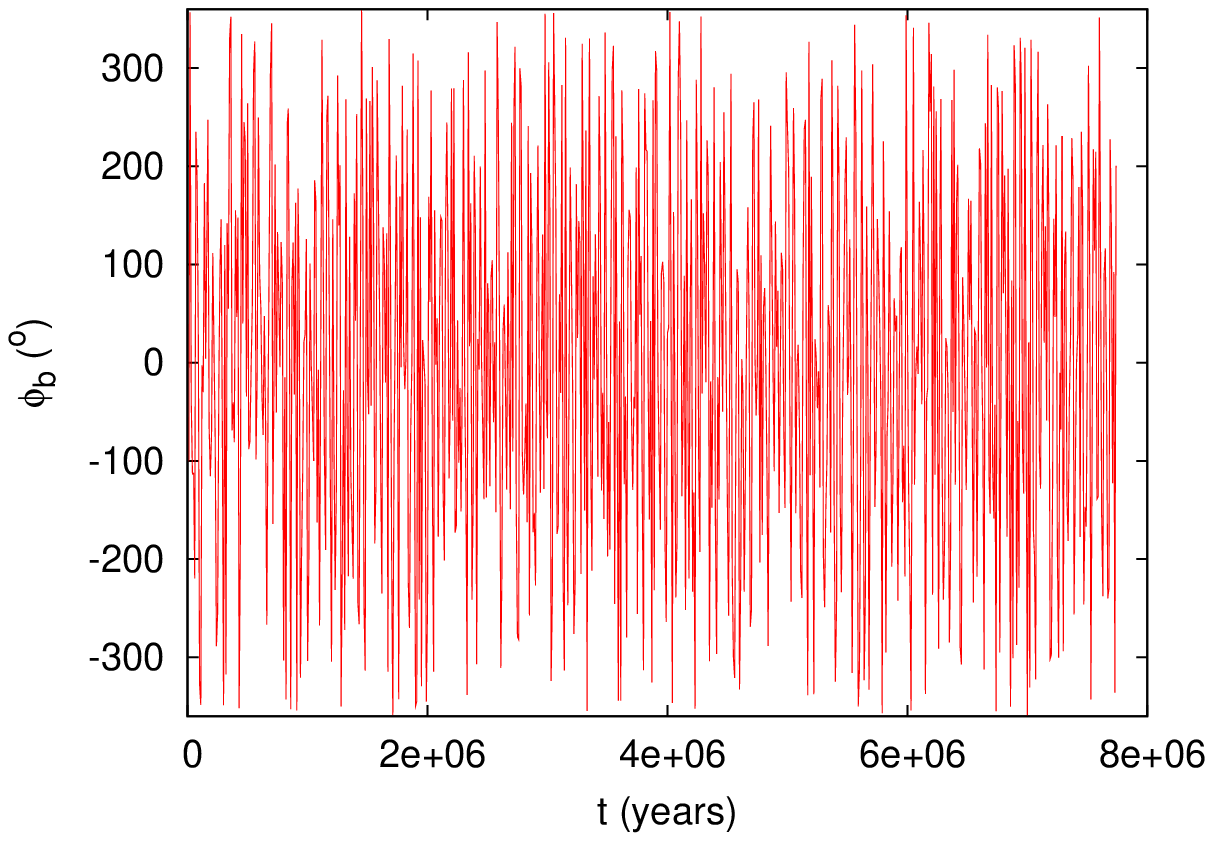}}
\caption{\small{Variation in the resonant argument $\phi_{b}$ given by Eq.~\ref{eq3} for the longest living test particle at 1.4 AU of model 4.}}
\label{fig:3} 
\end{subfigure}
\begin{subfigure}
{\includegraphics[width=3.5in]{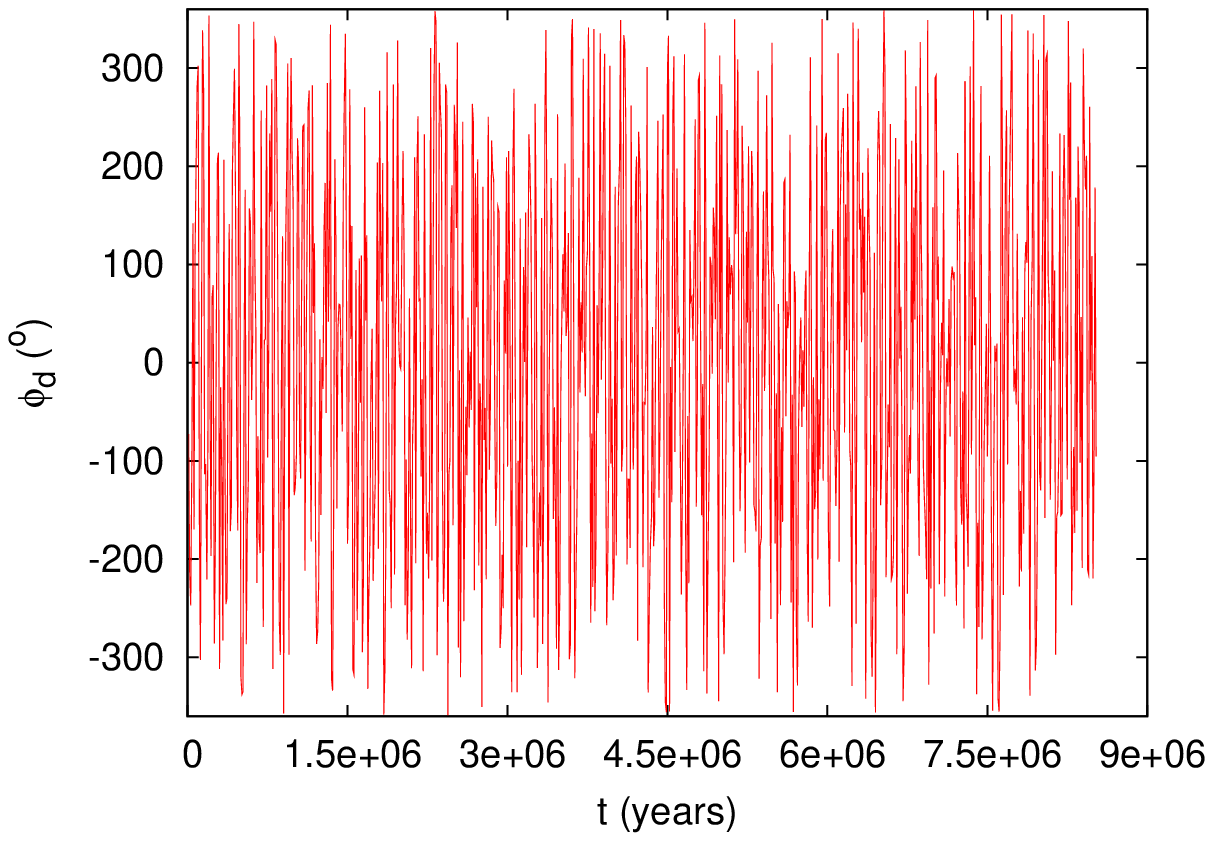}}
\caption{\small{Variation in the resonant argument $\phi_{d}$ given by Eq.~\ref{eq4} for the longest living test particle at 1.5 AU of model 4.}} 
\label{fig:4}
\end{subfigure}                
\end{figure} 
We therefore carried out a more detailed investigation in order to determine why test particles in these two regions displayed semi-stable behaviour across our entire suite of simulations.

One potential explanation for the behaviour is resonance trapping (e.g.~\cite{Lyk07}), and so we examined the potential mean motion resonances (MMR) located between 1.4 and 1.6 AU resulting from the two giant planets, HD~204313 b and d. Particles orbiting around $a\sim 1.5$~AU have a period $P_{TP_{1.5}}\sim$ 667 days, while particles orbiting around $a\sim 1.4$~AU have a period $P_{TP_{1.4}}\sim$ 601 days. Thus the ratio between the orbital periods of the particles trapped at 1.5~AU and HD~204313~d is $P_{d}/P_{TP_{1.5}} = 4.2 $, suggesting that test particles may possibly be trapped in a 1:4 MMR with HD~204313~d. Similarly, the ratio between the orbital periods of the particles trapped at 1.4~AU and HD~204313~b is $P_{b}/P_{TP_{1.4}} = 3.2 $, suggesting that test particles may possibly be trapped in a 1:3 MMR with HD~204313~b\footnote{We note that exact commensurability in the ratio of the periods is not required for the resonant angle to librate. Resonance can occur on a small “distance” from exact commensurability in dimensionless units of period ratio, which is called 'resonance width'~\cite{Mar13}.}. To test this scenario, we plotted the evolution of the resonant argument, $\phi$, for those MMRs:
\begin{equation}
\phi_{d}=\lambda_{TP}-4 \lambda_{d}+3 \omega_{TP}
\label{eq3}
\end{equation}
\begin{equation}
\phi_{b}=\lambda_{TP}-3 \lambda_{b}+2 \omega_{TP}
\label{eq4}
\end{equation}
where $\lambda$ is the mean longitude and $\omega$ the argument of the periastron. However we found that these resonant arguments did not exhibit libration or any periodic behaviour for any of the test particles tested -- as can be seen in Figures~\ref{fig:3} and~\ref{fig:4}.

Since the locations of the semi-stability regions do not match any strong MMRs with HD~204313~b or HD~204313~d, we examined whether the test particles might be trapped in regions where a number of overlapping weak and/or high order MMRs could be combining to create a region of "sticky chaos", acting to stabilise the test particles motion~\cite{Lyk07}. We calculated all possible resonances with HD~204313~b or HD~204313~d up to order 51 and plotted their individual resonant arguments with respect to both planets. However, we were unable to identify any clear resonant behaviour. We therefore conclude that no single weak/high order resonance is responsible for these semi-stable regions, although we note that test particles may be captured and stabilised as a result of the overlap between a number of these high-order resonances. In such a scenario, test particles would chaotically hop from one resonance to another, without spending any significant period of time trapped in any specific resonance. Such behaviour has been invoked in the past to explain regions of stability for other exoplanetary systems~\cite{Goz13}.

\begin{figure}[!t]
\centering
\includegraphics[width=7in]{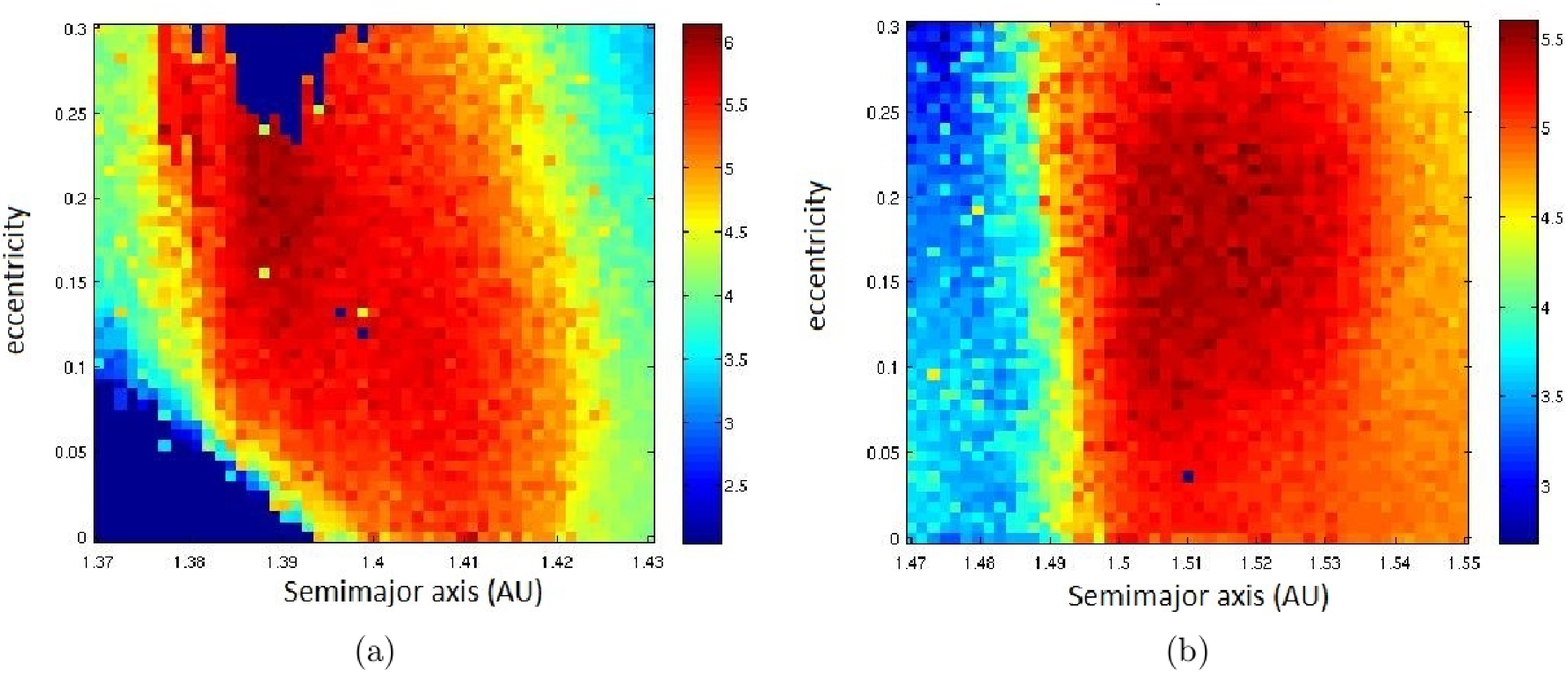}
\caption{\small{The mean dynamical lifetime of the test particles as a function of (a) the initial semimajor axis, $a$, and eccentricity, $e$, around 1.4 AU, and around (b) 1.5 AU. The lifetime in years is plotted with a logarithmic color scale.}}
\label{fig:5}
\end{figure}
In order to better define the two semi-stable regions, we produced high resolution maps of the mean dynamical test particle lifetime as a function of their initial orbital elements by running additional simulations with $10^{6}$ test particles distributed around 1.4~AU and 1.5~AU. Figure~\ref{fig:5} shows the mean dynamical lifetime of the test particles as a function of the initial semi-major axis, and eccentricity. Whilst stable zones are observed for orbits spanning the entire range of orbital eccentricities tested (between 0 and 0.3), the stability of the test particles is clearly most strongly influenced by the semi-major axis of the test particles orbit. We thus conclude that fine tuning of initial semi-major axis is required to obtain stable orbits within the HZ. However, while all simulations were run for $10^{8}$ years, as can be seen in Figure~\ref{fig:5}, the simulations were stable for no longer than $10^{6}$ years.

\begin{figure}[!t]
\centering
\includegraphics[width=3.5in]{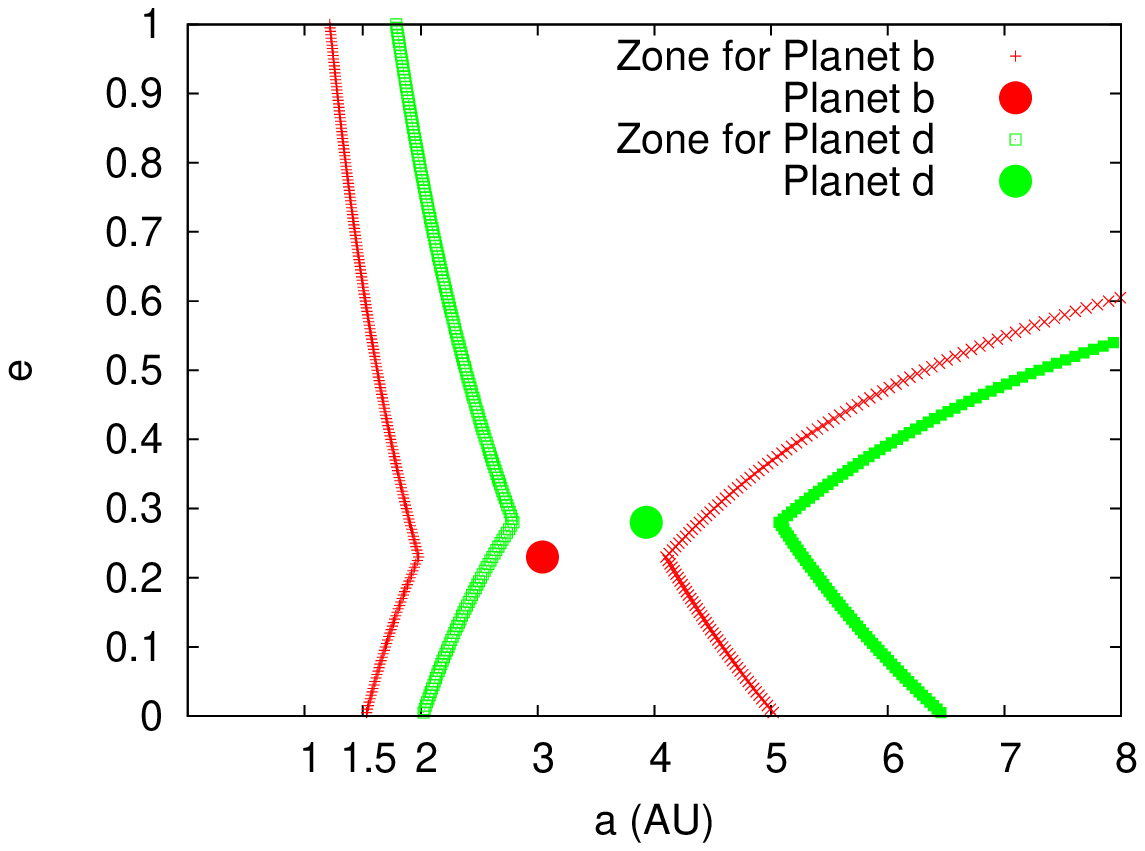}                  
\caption{\small{Theoretical stability map for  HD~204313. The dots represent the location of the planets HD~204313 b (red) and HD~204313 d (green), while the dashed lines delimit the extended orbit crossing regions around HD~204313 b (red) and HD~204313 d (green).}} 
\label{fig:6}  
\end{figure}
We now compare the location of the two semi-stable regions inside the HZ with the stability map analytically derived using the criterion of~\cite{Giu13} -- see Figure~\ref{fig:6}. The first remark is that planets b and d are mutually located inside their unstable regions and our work therefore supports the conclusions of~\cite{Rob12} that the 3:2 MMR is responsible for maintaining these two planets in a stable configuration. Whilst the inner limit of the unstable zone of HD~204313 d (i.e. to the left of the innermost green curve in Figure~\ref{fig:6}) is located between 1.8~AU for highly eccentric particles and 2.8~AU for particles with $e \sim 0.3$, the inner limit of the unstable zone of HD~204313 b (i.e. to the left of the innermost red curve in Figure~\ref{fig:6}) is between 1.4~AU for highly eccentric particles and 1.8~AU for particles with $e \sim 0.25$. This map therefore explains why no stable orbits are found exterior to 1.5 AU within the HZ - beyond this semi-major axis, the proximity between additional bodies and the known planets would result in the disruption of the system.
 
\section*{Conclusions}
We propose a new method to investigate the stability of potential terrestrial planets in the habitable zone of multiplanetary systems. This method combines numerical integrations and an analytical criterion to assess the suitability of the HZ to host additional long-term stable objects. Using atmospheric model criterion, we first defined the location of the HZ before testing its appropriateness to host massless test particles by numerically integrating their orbits on a timescale required for the establishment of life. We examined the resulting stable regions in terms of the presence of any resonance mechanisms, and produced high resolution maps of the test particles lifetime as a function of their initial orbital elements inside the identified stable regions. As a final step, we compared the location of the numerical stable zones with the region allowed by an analytical criteria which checks for orbit crossings. This approach provides a more complete picture of the dynamics of the HZ. 
We applied this method to the system HD~204313, which is mainly composed of a Sun-like star and two giant planets, HD~204313~b and d, orbiting respectively at 3.04 and 3.95~AU (we ignored HD~204313~c). We report the following results and conclusions:
\begin{itemize}
\item Using the `Runaway greenhouse effect' and `Maximum greenhouse effect', we defined the HZ of HD~204313~b and d as the region between 1.1--1.9~AU.
\item By distributing massless test particles throughout the HZ of the HD~204313 system and testing their stability with numerical simulations, we found two semi-stable regions near 1.4 and 1.5 AU. 
\item Although no single mean motion resonance up to an order 51 was identified as controlling the evolution of the test particles in those regions, our investigations do not exclude that particles could be trapped due to overlapping weak and high-order resonances with the two outer planets of the system.
\item Using the analytical criterion from~\cite{Giu13}, we confirm that the 3:2 MMR between HD~204313~b and d is required to maintain this system in a stable configuration. Moreover, if not protected by any resonance mechanism, no additional planets can be located in the HZ with $a >$1.5~AU at the risk of destabilizing the orbits of the two outer giant planets.
\item However, it must be noted that these two semi-stable regions remain stable for only a short period of $\sim~9\times10^{6}$ years, which is less than the timescale required for the emergence of life ($\sim 10^{9}$~years). Thus those two semi-stable regions are not suitable for a terrestrial planet to develop life.
\end{itemize}
While we did not find a zone of potential long-term stability and habitability for planets in the HZ of the HD~204313 system, this study established a framework for a larger project that will study many different systems using a similar method. One could follow this approach for all known multiple systems with well constrained orbital elements. Generally one expects low eccentricity systems would have a higher chance of hosting stable Earth-like planet in their habitable zone. If we find multiple systems which can host stable Earth-mass planets in their HZ, these systems could be targeted to search for low-mass planets in future surveys.

\section*{Acknowledgments}
All simulations were run on the Swinburne supercomputer. LJ was supported by the Swinburne Centre for Astrophysics \& Supercomputing. We thank Rosemary Mardling for useful discussions, and the referees for their feedback.


\begin{thebibliography}{1}
\normalsize{}
\bibitem{Bor13}
Borucki, W.~J. et al., ``Kepler-62: A Five-Planet System with Planets of 1.4 and 1.6 Earth Radii in the Habitable Zone'', \emph{Science}, 2013, Vol. 340, pages 587-590

\bibitem{Kas93}
Kasting, J.~F. and Whitmire, D.~P. and Reynolds, R.~T., ``Habitable Zones around Main Sequence Stars'', \emph{ICARUS}, 1993, Vol. 101, pages 108-128 

\bibitem{Chy93}
Chyba, C.~F., ``The violent environment of the origin of life: Progress and uncertainties'', \emph{Geochimica et Cosmochimica Acta}, 1993, Vol. 57, pages 3351-3358

\bibitem{Qia11}
Qian, S.-B. and Liu, L. and Liao, W.-P. and Li, L.-J. and Zhu, L.-Y. and Dai, Z.-B. and He, J.-J. and Zhao, E.-G. and Zhang, J. and Li, K., ``Detection of a planetary system orbiting the eclipsing polar HU Aqr'', \emph{Monthly Notices of the Royal Astronomy Society}, 2011, Vol. 414, pages L16-L20

\bibitem{Hor11}
Horner, J. and Marshall, J.~P. and Wittenmyer, R.~A. and Tinney, C.~G., ``A dynamical analysis of the proposed HU Aquarii planetary system'', \emph{Monthly Notices of the Royal Astronomy Society}, 2011, Vol. 416, pages L11-L15

\bibitem{Wit12}
Wittenmyer, R.~A. and Horner, J. and Marshall, J.~P. and Butters, O.~W. and Tinney, C.~G., ``Revisiting the proposed planetary system orbiting the eclipsing polar HU Aquarii'', \emph{Monthly Notices of the Royal Astronomy Society}, 2012, Vol. 419, pages 3258-3267

\bibitem{Hin08}
Hinse, T.~C. and Michelsen, R. and J{\o}rgensen, U.~G. and Go{\'z}dziewski, K. and Mikkola, S., ``Dynamics and stability of telluric planets within the habitable zone of extrasolar planetary systems. Numerical simulations of test particles within the HD 4208 and HD 70642 systems'', \emph{Astronomy \& Astrophysics}, 2008, Vol. 488, pages 1133-1147

\bibitem{San07}
S{\'a}ndor, Z. and S{\"u}li, {\'A}. and {\'E}rdi, B. and Pilat-Lohinger, E. and Dvorak, R., ``A stability catalogue of the habitable zones in extrasolar planetary systems'', \emph{Monthly Notices of the Royal Astronomy Society}, 2007, Vol. 375, pages 1495-1502

\bibitem{Giu13}
Giuppone, C.~A. and Morais, M.~H.~M., and Correia, A.~C.~M., ``A semi-empirical stability criterion for real planetary systems with eccentric orbits '', \emph{arXiv}: 1309.6861, 2013

\bibitem{Dec13}
Deck, K.~M. and Payne, M. and Holman, M.~J., ``First-order Resonance Overlap and the Stability of Close Two-planet Systems'', \emph{The Astrophysical Journal}, 2013, Vol. 774, page 129

\bibitem{Lev94}
Levison, H.~F. and Duncan, M.~J., ``The long-term dynamical behavior of short-period comets'', \emph{ICARUS}, Vol. 108, 1994, pages 18-36

\bibitem{Kha09}
Kharchenko, N.~V. and Roeser, S., ``All-sky Compiled Catalogue of 2.5 million stars'', \emph{VizieR Online Data Catalog}, 2009, Vol. 1280

\bibitem{Cas11}
Casagrande, L. and Sch{\"o}nrich, R. and Asplund, M. and Cassisi, S. and Ram{\'{\i}}rez, I. and Mel{\'e}ndez, J. and Bensby, T. and Feltzing, S., ``New constraints on the chemical evolution of the solar neighbourhood and Galactic disc(s). Improved astrophysical parameters for the Geneva-Copenhagen Survey'', \emph{Astronomy \& Astrophysics}, 2011, Vol. 530, pages A138

\bibitem{Rob12} 
Robertson, P., Horner, J. Wittenmyer, R. A., Endl, M., Cochran, W. D., MacQueen, P.J., Brugamyer, E. J., Simon, A. E., Barnes. S. I. and Caldwell, C., ``A Second Giant Planet in 3:2 Mean-motion Resonance in the HD 204313 system'', \emph{The Astrophysical Journal}, Vol. 754, Issue 1, 2012, article id. 50

\bibitem{Sne73}
Sneden, C., ``The nitrogen abundance of the very metal-poor star HD 122563'', \emph{The Astrophysical Journal}, 1973, Vol. 184, pages 839-849 

\bibitem{Seg10}
S{\'e}gransan, D. and Udry, S. and Mayor, M. and Naef, D. and Pepe, F. and Queloz, D. and Santos, N.~C. and Demory, B.-O. and Figueira, P. and Gillon, M. and Marmier, M. and M{\'e}gevand, D. and Sosnowska, D. and Tamuz, O. and Triaud, A.~H.~M.~J, ``The CORALIE survey for southern extrasolar planets. XVI. Discovery of a planetary system around HD 147018 and of two long period and massive planets orbiting HD 171238 and HD 204313'', \emph{Astronomy \& Astrophysics}, 2010, Vol. 511, page A45

\bibitem{Mayo11}
Mayor, M. and Marmier, M. and Lovis, C. and Udry, S. and S{\'e}gransan, D. and Pepe, F. and Benz, W. and Bertaux, J.~-. and Bouchy, F. and Dumusque, X. and Lo Curto, G. and Mordasini, C. and Queloz, D. and Santos, N.~C.., ``The HARPS search for southern extra-solar planets XXXIV. Occurrence, mass distribution and orbital properties of super-Earths and Neptune-mass planets'', \emph{arXiv}: 1109.2497v1, 2011

\bibitem{Und03}
Underwood, D.~R. and Jones, B.~W. and Sleep, P.~N., ``The evolution of habitable zones during stellar lifetimes and its implications on the search for extraterrestrial life'', \emph{International Journal of Astrobiology}, Vol. 2, 2003, pages 289-299

\bibitem{Jon05}
Jones, B.~W. and Underwood, D.~R. and Sleep, P.~N., ``Prospects for Habitable Earths in Known Exoplanetary Systems '', \emph{The Astrophysical Journal}, Vol. 622, 2005, pages 1091-1101

\bibitem{Jon06}
Jones, B.~W. and Underwood, D.~R.and Sleep, P.~N., ``Which exoplanetary systems could harbour habitable planets?'', \emph{International Journal of Astrobiology}, 2006, Vol. 5, pages 251-259

\bibitem{Men03}
Menou, K. and Tabachnik, S. ,``Dynamical Habitability of Known Extrasolar Planetary Systems'', \emph{The Astrophysical Journal}, 2003, Vol. 583, pages 473-488

\bibitem{Kop13}
Kopparapu, R.~K. and Ramirez, R. and Kasting, J.~F. and Eymet, V. and Robinson, T.~D. and Mahadevan, S. and Terrien, R.~C. and Domagal-Goldman, S. and Meadows, V. and Deshpande, R., ``Habitable Zones around Main-sequence Stars: New Estimates'', \emph{The Astrophysical Journal}, Vol. 770, 2013, page 82

\bibitem{Cham99}
Chambers, J.~E.,``A hybrid symplectic integrator that permits close encounters between massive bodies'', \emph{Monthly Notices of the Royal Astronomy Society}, Vol. 304, 1999, pages 793-799

\bibitem{Wis80}
Wisdom, J., ``The resonance overlap criterion and the onset of stochastic behavior in the restricted three-body problem'', \emph{Astronomical Journal}, 1980, Vol. 85, pages 1122-1133

\bibitem{Lyk07}
Lykawka, P.~S. and Mukai, T., ``Resonance sticking in the scattered disk'', \emph{ICARUS}, 2007, Vol. 192, pages 238-247

\bibitem{Mar13}
Mardling, R.~A., ``New developments for modern celestial mechanics - I. General coplanar three-body systems. Application to exoplanets'', \emph{Monthly Notices of the Royal Astronomy Society}, 2013, Vol. 435, pages 2187-2226

\bibitem{Goz13}
Gozdziewski, K. and Slonina, M. and Migaszewski, C. and Rozenkiewicz, A., ``Testing a hypothesis of the $\nu$ Octantis planetary system'', \emph{Monthly Notices of the Royal Astronomy Society}, 2013, Vol. 430, pages 533-545

\end{thebibliography}
\end{document}